\documentclass[nofootinbib,superscriptaddress,a4paper, twocolumn, floatfix]{revtex4-2}
\usepackage[english]{babel}
\usepackage[utf8]{inputenc}
\usepackage[colorinlistoftodos, color=green!40, prependcaption]{todonotes}
\usepackage{amsthm}
\usepackage{mathtools}
\usepackage{physics}
\usepackage{braket}
\usepackage{listings}
\usepackage{xcolor}
\usepackage{graphicx}
\usepackage[left=18.5mm,right=15mm,top=30mm,columnsep=15pt]{geometry} 
\usepackage{adjustbox}
\usepackage{placeins}
\usepackage[T1]{fontenc}
\usepackage{lipsum}
\usepackage{csquotes}
\usepackage[pdftex, pdftitle={Article}, pdfauthor={Author}]{hyperref} 
\usepackage{booktabs}
\usepackage{amssymb}
\usepackage{bm}
\usepackage[ruled,vlined]{algorithm2e}
\usepackage[caption=false]{subfig}

\begin{document}

\title{Stabilizer configuration interaction: Finding molecular subspaces with error detection properties}

\author{Abhinav Anand}
\email[E-mail:]{abhinav.anand@duke.edu}
\affiliation{Duke Quantum Center, Duke University, Durham, NC 27701, USA.}
\affiliation{Department of Electrical and Computer Engineering, Duke University, Durham, NC 27708, USA.}

\author{Kenneth R. Brown}
\email[E-mail:]{kenneth.r.brown@duke.edu}
\affiliation{Duke Quantum Center, Duke University, Durham, NC 27701, USA.}
\affiliation{Department of Electrical and Computer Engineering, Duke University, Durham, NC 27708, USA.}
\affiliation{Department of Physics, Duke University, Durham, NC 27708, USA.}
\affiliation{Department of Chemistry, Duke University, Durham, NC 27708, USA.}

\date{\today}

\begin{abstract}
    In this work, we explore a new approach to designing both algorithms and error detection codes for preparing approximate ground states of molecules. We propose a classical algorithm to find the optimal stabilizer state by using excitations of the Hartree-Fock state, followed by constructing quantum error-detection codes based on this stabilizer state using codeword-stabilized codes. Through various numerical experiments, we confirm that our method finds the best stabilizer approximations to the true ground states of molecules up to 36 qubits in size. Additionally, we construct generalized stabilizer states that offer a better approximation to the true ground states. Furthermore, for a simple noise model, we demonstrate that both the stabilizer and (some) generalized stabilizer states can be prepared with higher fidelity using the error-detection codes we construct. Our work represents a promising step toward designing algorithms for early fault-tolerant quantum computation.
\end{abstract}

\maketitle

\section{Introduction}
In recent years, various demonstrations of fault-tolerant (FT) operations on physical devices~\cite{google2023suppressing, krinner2022realizing,andersen2020repeated, bluvstein2024logical, ryan2021realization, egan2021fault, postler2022demonstration} have been carried out, leading to proposals for their use in early fault-tolerant~\cite{katabarwa2024early, campbell2021early} quantum computing. 
A common theme of these proposals is developing methods~\cite{lin2022heisenberg, ding2023even, zhang2022computing, kliuchnikov2023shorter, vandaele2024lower, kissinger2019reducing} that require low overhead (such as the number of logical qubits, operations, and T-count) for the fault-tolerant implementation of different algorithms. 
While this is an important and promising research direction, the overhead~\cite{nelson2024assessment} associated with these proposals can still be prohibitively high.

A key reason for the extremely high overhead is the significant resources required for encoding information in logical qubits.
To address this, we explore a new approach that designs both algorithms and error detection codes for preparing approximate ground states of molecules~\cite{aspuru2005simulated, abrams1997simulation, Anand2021Quantum}. 
Our approach involves using the approximate eigenspace (spanned by a few stabilizer states) of the physical system we are simulating as the code space of the error detection codes.

Recent studies~\cite{peruzzo2014variational, motta2020determining, mcardle2019variational, cerezo2020variational, bharti2022noisy} have proposed various methods for constructing approximate eigenstates of different physical systems. 
However, most of these methods involve building parameterized quantum circuits~\cite{anand2023hamiltonians, anand2023leveraging, wecker2015progress,grimsley2019adaptive,ryabinkin2018qubit,ryabinkin2020iterative,tang2021qubit,McClean2016theoryofvqe} or Clifford circuits~\cite{kottmann2022optimized,schleich2023partitioning, anand2022quantum, ravi2022cafqa, wang2024demonstration} and performing optimization~\cite{schuld2019evaluating, mitarai2018quantum, kottmann2021feasible, wierichs2022general, gacon2021simultaneous, anand2021natural, bonet2023performance,sweke2020stochastic} to find the best approximations.
This makes these methods computationally expensive and limits their applicability to smaller systems as they have various limitations.~\cite{mcclean2018barren, cerezo2021cost, wang2021noise, marrero2021entanglement}
To overcome this limitation, we propose an efficient classical algorithm for constructing stabilizer approximations of the ground states~\cite{li2022stabilizer, wang2023stabilizer, sun2024stabilizer} of various molecules.
We present results from various numerical experiments that approximate the ground state of various molecules of size up to 36 qubits.
Our numerical experiments range from taking just seconds for smaller molecules like H$_2$ to several hours for larger ones like Cr$_2$, which is a significant improvement over previous methods, where computations of this scale could take anywhere from minutes to weeks~\cite{ravi2022cafqa}.

Subsequently, we construct quantum codes using codeword stabilized codes~\cite{cross2008codeword} based on the stabilizer approximations of the ground states. 
These codes are error detection codes with a single logical qubit and require minimal resources to prepare the states using error-detection and post-selection.

While the states we construct may not achieve the desired accuracy for most of the molecules considered in this study, our work represents a significant step towards the design of algorithms suitable for implementation on early fault-tolerant quantum computers.

The remainder of the paper is organized as follows:
We begin with a review of preliminary theory and background information in Sec.~\ref{section:prelims}.
The algorithm and results from various numerical experiments are discussed in Sec.~\ref{sec:method}. 
In Sec.~\ref{sec:codes}, we present the details of the codes and demonstrate the error-detection property of the method.
Finally, we provide concluding remarks in Sec.~\ref{sec:conclusion}.

\section{Theory and Background}\label{section:prelims}
In this section, we review some of the essential theory and background information required for the rest of the paper. 

\subsection{Hartree-Fock and configuration interaction wavefunctions}\label{subsec:CI}
In this article, we consider the second quantized formalism, where the state of a qubit represents the occupancy of a spin-orbital basis function.

The Hartree-Fock (HF) state~\cite{szabo2012modern,helgaker2013molecular} in this formalism is a single product wavefunction with the minimal energy.
One solves a set of self-consistent field (SCF) equations to determine the optimal molecular orbitals that minimize the total electronic energy of the system.
Using the optimized molecular orbitals as a basis, the $n$-qubit Hartree-Fock state can be written as:
\begin{equation}
\ket{\text{HF}} = \ket{\underbrace{11....11}_{n_{\text{o}}}00....00},
\end{equation}
where $n_{o}$ is the number of occupied spin orbitals.

Using the HF state as a reference state, one can create a linear combination of configurations (excitations).
The exact ground state solution is the full configuration interaction (FCI) wavefunction~\cite{szabo2012modern,helgaker2013molecular} $\ket{\text{FCI}}$ and requires a sum over all possible excitations.
\begin{align}
    \ket{\text{FCI}} &= \sum_k^{n_{\text{o}}}\hat{C}_k \ket{\text{HF}}, \\
    \hat{C}_k &= \sum_{i<j<..}^{\text{o}} \sum_{a<b<..}^{\text{vir}} c_{ij...}^{ab...} \hat{a}_{a}^{\dagger} \hat{a}_{b}^{\dagger} .... \hat{a}_i \hat{a}_j ...., 
\end{align}
where $c_{ij...}^{ab...}$ denotes complex coefficients and $\hat{C}_k$ is an operator consisting of $k$ annihilation operators, $\hat{a}_i$,  and $k$ creation operators, $\hat{a}_{a}^{\dagger}$, acting on the occupied and virtual orbitals, respectively. 
The operator $\hat{C}_k$ generates a $k$-fold configuration.

\subsection{Stabilizer states}\label{sec:stab_state}
A stabilizer state~\cite{gottesman1997stabilizer} is defined as a $n$-qubit pure state $\ket{\psi_s}$, which is stabilized by an abelian subgroup $\mathcal{S}$ of the Pauli group $\mathcal{G}_n$, i.e.,
\begin{equation}
    P\ket{\psi_s} = \ket{\psi_s}; \medskip \forall P \in \mathcal{S},
\end{equation}
where $P$ is a Pauli-string (tensor product of Pauli matrices $I$, $X$, $Y$ and $Z$) on $n$-qubits.

The stabilizer state $\ket{\psi_s}$, is local-Clifford equivalent to a graph state~\cite{Van_den_Nest_2004, zeng2007local}, so there exists a local Clifford unitary that maps every stabilizer to the form $X_{v}Z_{\{n\}}$, where $v$ represents a vertex in the graph and $\{n\}$ is the corresponding set of neighbors.
 We refer to this form ($X_{v}Z_{\{n\}}$) of the stabilizers as standard form.

It is well known~\cite{garcia2017geometry} that given two $n$-qubit stabilizer states, $\ket{\psi}$ and $\ket{\phi}$, $\frac{\ket{\psi} + i^{l}\ket{\phi}}{\sqrt{2}}$ is also a stabilizer state, iff $\langle \phi \ket{\psi} = 0$ and $\ket{\phi}= P\ket{\psi}$, where $l \in \{0,1,2,3\}$ and $P \in \mathcal{G}_n$.

\subsection{Codeword stabilized (CWS) code}
A $[[n,K]]$ codeword stabilized code~\cite{cross2008codeword} is defined by a stabilizer group $\mathcal{S}$ and a set of $n$-qubit Pauli-strings, $\mathcal{W}=\{w_k\}_{k=1}^K$, called the word operators. 
The first word operator $w_1$ is always chosen to be the identity operator.
The code is then spanned by the basis vectors of the form
\begin{equation}
    \ket{\psi_{w_{k}}} = w_k\ket{\psi_s},
\end{equation}
where $\ket{\psi_s}$ is the corresponding stabilizer state.

If the stabilizers of the code are in the standard form $X_{v}Z_{\{n\}}$, then one can transform the word operators to the standard form, strings of $Z$s.
Additionally, any single qubit error acting on the codewords $\{\ket{\psi_{w_{k}}}\}$ is equivalent to another (possibly multi-qubit) error consisting only of $Z$s.
Since all errors become $Z$s, one can treat this as a classical error model and find the set of errors that can be detected by the CWS code (see Theorem 3 of Ref.~\cite{cross2008codeword}).

The CWS code is a $[[n,k=\log_2K]]$ stabilizer code (see Theorem 5 of Ref.~\cite{cross2008codeword}) if the word operators $\mathcal{W}$ form an abelian group.
It should be noted that in general the word operators do not form a group.

\section{Method and Results}\label{sec:method}
In this section, we present the details of the classical algorithm to find the best classical state to approximate ground states and present results from different numerical simulations.

\subsection{Stabilizer Configuration Interaction}\label{sec:stab_CI}
Inspired by the full configuration interaction (FCI) method, we propose a method to generate stabilizer approximations of ground states by a linear superposition of configurations (excitations).
We now describe the method in detail.

\subsubsection{Reference state}\label{subsec:refst}
The starting point of our method is choosing a reference state, from which subsequent excitations can be generated. 
The Hartree-Fock (HF) state~(Sec. \ref{subsec:CI}) is chosen as the reference state in our method. 
It is the best product state approximation to the ground state of the system, as it captures the essential mean-field characteristics of the electronic structure. 
The HF state, being a stabilizer state, allows us to generate and manage excitations systematically.

\subsubsection{Excitations}\label{subsec:exc}
Next, we introduce different excitations in the reference state to create a state that capture the effects of electron correlation (similar to FCI wavefunction~(\ref{subsec:CI})) and improve the stabilizer approximation. 
We generate these excitations carefully to preserve the stabilizer nature of the resulting state while also maintaining the symmetry of the physical system.

We select a set of operators $\{E_1, E_2, E_3 ...\}$ to include excitations in the reference state iteratively, as:
\begin{equation}\label{eq:iter_stab}
    \ket{\psi_{i+1}} = \frac{(\mathcal{I} + (-1)^l E_{i+1})\ket{\psi_i}  }{\sqrt{2}},
\end{equation}
where $l \in \{ 0, 1\}$ and $\ket{\psi_0} = \ket{\text{HF}}$.
The final state $\ket{\psi_{n}}$ generated by a set of operators $\{E_1, E_2, ... , E_n\}$ is a stabilizer state with the same symmetry as the reference state if the following conditions are satisfied:
\begin{enumerate}
    
    \item Any operator, $E_i$, acts on equal number of occupied and unoccupied orbitals and excites particles to the same spin orbitals and is of the form $X_{\{o\}}X_{\{u\}}$, where $\{o\}$ and $\{u\}$ represent the set of occupied and unoccupied spin orbitals, respectively.
    \item The number of independent single excitation generators $\{X_{o_{1}}X_{u_{1}}$, ...,  $X_{o_{m}}X_{u_{m}}\}$ is less than or equal to the number of particles, i.e., $m \le n_e$, where $o_i$ and $u_i$ are orbitals with the same spin symmetry.
    Further, any two independent single excitation generators $X_{o_{i}}X_{u_{i}}$ and $X_{o_{j}}X_{u_{j}}$ are fully disjoint, i.e., $X_{o_{i}}X_{u_{i}} \cap X_{o_{j}}X_{u_{js}} = \emptyset$.
\end{enumerate}

One can check that if the above conditions are met, that $\bra{\psi_{i-1}}E\ket{\psi_{i-1}} = 0$, thus the new state $\ket{\psi_{i+1}}$ (Eq.~\ref{eq:iter_stab}) is a stabilizer state (see Sec.~\ref{sec:stab_state}).
So, if we select the reference state to be a stabilizer state, the final state will also be a stabilizer state.

It should be noted that using such a set of operators one can generate a maximum of $2^{n_o}$ excitations, which is much lower than the total number of excitations, $2^{n_o + n_u}$, as $n_u >> n_o$ in many practically relevant cases.

The procedure to generate a set of valid excitation generators is summarized below.
First, we construct two matrices corresponding to the two spin excitation generators ($\alpha$ and $\beta$), each of size $n_o/2 \times n_u/2$, where $n_o$ and $n_u$ are the number of occupied and  virtual (unoccupied) orbitals, respectively.
Every element of the matrix holds the corresponding orbital indices, ($o_i, u_j$), such that the action of the operator $X_{o_{i}}X_{u_{j}}$ on the reference state generates the corresponding single excitation.

We then sample $n_o$ elements $\{ (o_{i_{1}}, u_{j_{1}})$, $(o_{i_{2}}, u_{j_{2}})$,..., $(o_{i_{n_{o}}}, u_{j_{n_{o}}})\}$ from the two matrices, such that no two entries belong to the same row or column.
A set of excitation generators is constructed using the single excitation generators $X_{o_{i}}X_{u_{j}}$ corresponding to each of the sampled elements $(o_i, u_j)$.
The number of different excitation generators in the set can vary from 1 to $n_o$.

\subsubsection{Algorithm}\label{sec:alg_SCI}
Given a molecular Hamiltonian in the form $H=\sum h_i P_i$,  obtained via a fermion-to-qubit transformation~\cite{bravyi2002fermionic, jordan1928pauli} of the second-quantized Hamiltonian,  and the corresponding Hartree-Fock state $\ket{\psi_0}$.
We now describe the procedure for finding stabilizer approximations to ground states of the molecule:
\begin{enumerate}
    \item Generate all valid sets of excitation generators by following the procedure mentioned above.
    \item For every set $\{E_1, E_2, E_3 ...\}$ generate stabilizer states iteratively using the Hartree-Fock state as the reference state, following the Eq.~\ref{eq:iter_stab}. 
    \item Calculate the ground state energy corresponding to the molecular Hamiltonian and output the state, $\ket{\psi_n}$, that has the lowest energy.
\end{enumerate}

The above algorithm looks at all the stabilizer states with real amplitudes and only outputs the state with the lowest energy.
While energy is the primary metric considered here, one can choose other metrics (similarity, overlap, etc.) and find the best stabilizer state that satisfies the chosen criterion.

\subsection{Adaptive stabilizer CI}\label{sec:ad_stab_CI}
The number of valid excitation generator sets grows combinatorially with both the number of occupied orbitals, $n_o$, and number of unoccupied  orbitals, $n_u$.

Consequently, the runtime of the algorithm scales approximately as $\mathcal{O}({n}^{n_o})$, where $n = n_u + n_o$. 
While this is manageable for small molecules, it can become computationally impractical as the system size increases. 
Taking inspiration from adaptive methods, such as the ones proposed in Refs.~\cite{grimsley2019adaptive,ryabinkin2018qubit,ryabinkin2020iterative,tang2021qubit}, we propose a method that adaptively includes excitations of higher order and overcomes the scaling problem.

The modified algorithm is described in detail below.
\begin{enumerate}
    \item 
    Construct the set of all valid double excitation generators of the form  $ X_{o}X_{u} $, where $o = \{o_i, o_j \}$ and $u = \{u_i, u_j \}$ represent sets of occupied and unoccupied orbitals, respectively. Each pair ($o_k, u_k$) must consist of orbitals with the same spin, with one occupied and the other unoccupied.
    \item Generate all the stabilizer state of the form:
    \begin{align}\label{eq:ad_stab}
        \ket{\psi_{i+1}}&= \frac{ (\mathcal{I} + (-1)^l X_{o} X_{u})\ket{\psi_i}  }{\sqrt{2}}, \text{ and} \nonumber \\
        \ket{\psi_{i+1}^{'}} &= \frac{ (\mathcal{I} + (-1)^l X_{o} X_{u} E_{i-1})\ket{\psi_{i-1}}  }{\sqrt{2}},  \text{if } i>0 
    \end{align} 
    where $l\in\{0,1\}$ and $E_{i-1}$ is the excitation operator used to generate $\ket{\psi_i}$.
    \item Select the state(s) $\ket{\psi_{i+1}}$ with the lowest energy and remove the orbitals $\{o, u\}$ involved in the excitation included in $\ket{\psi_{i+1}}$ from consideration in the next steps. 
    \item Repeat steps 1-3, until all particles have been considered for an excitation.
\end{enumerate}

The above procedure scales as $\mathcal{O}(n^3)$, where $n=n_o + n_u$ is the total number of qubits and allows for investigating larger systems.
This scaling arises because the number of valid excitation generators in step 1 grows approximately as $\mathcal{O}(n^2)$. 
Consequently, the number of stabilizer states that need to be checked in step 2 also scales as $\mathcal{O}(n^2)$.
Since steps 1–3 are repeated $n_o$ times, the total runtime scales approximately as $\mathcal{O}(n^3)$.

It should be noted that, while the adaptive stabilizer CI algorithm exhibits favorable scaling, it only succeeds in finding the optimal stabilizer state ($\ket{\psi_n}$, as described in Sec.~\ref{sec:alg_SCI}) if there exists an energetically favorable path from the reference state to that state, i.e., one can reach $\ket{\psi_n}$ from $\ket{\psi_0}$ by repeatedly applying Step 2.
This limitation arises because the algorithm only accepts new states with additional excitations if they lead to a decrease in energy. 
To enhance exploration of the different configurations, the condition in Step 3 can be modified, for example, by retaining a fixed number of low-energy states rather than only the lowest energy state or allowing both singles and double excitations, thus allowing for broader exploration of configurations.
We leave such explorations to future studies.

\subsection{State preparation circuits}
In this section, we present the steps to construct the stabilizer states discussed in the previous sections.
It should be noted that there exists a Clifford circuit, $\mathcal{C}$, such that any stabilizer state $\ket{\psi_s}$ can be constructed by the action of $\mathcal{C}$ on $\ket{0}^{\bigotimes n}$, i.e., $\ket{\psi_s} = \mathcal{C}\ket{0}^{\bigotimes n}$.
However, this circuit can be quite complex, so we propose an alternative construction based on the Hadamard test protocol~\cite{cleve1998quantum}, that uses additional ancilla qubits and measurement to prepare the desired state.

\begin{figure}[htbp!]
    \centering
    \includegraphics[width=0.65\linewidth]{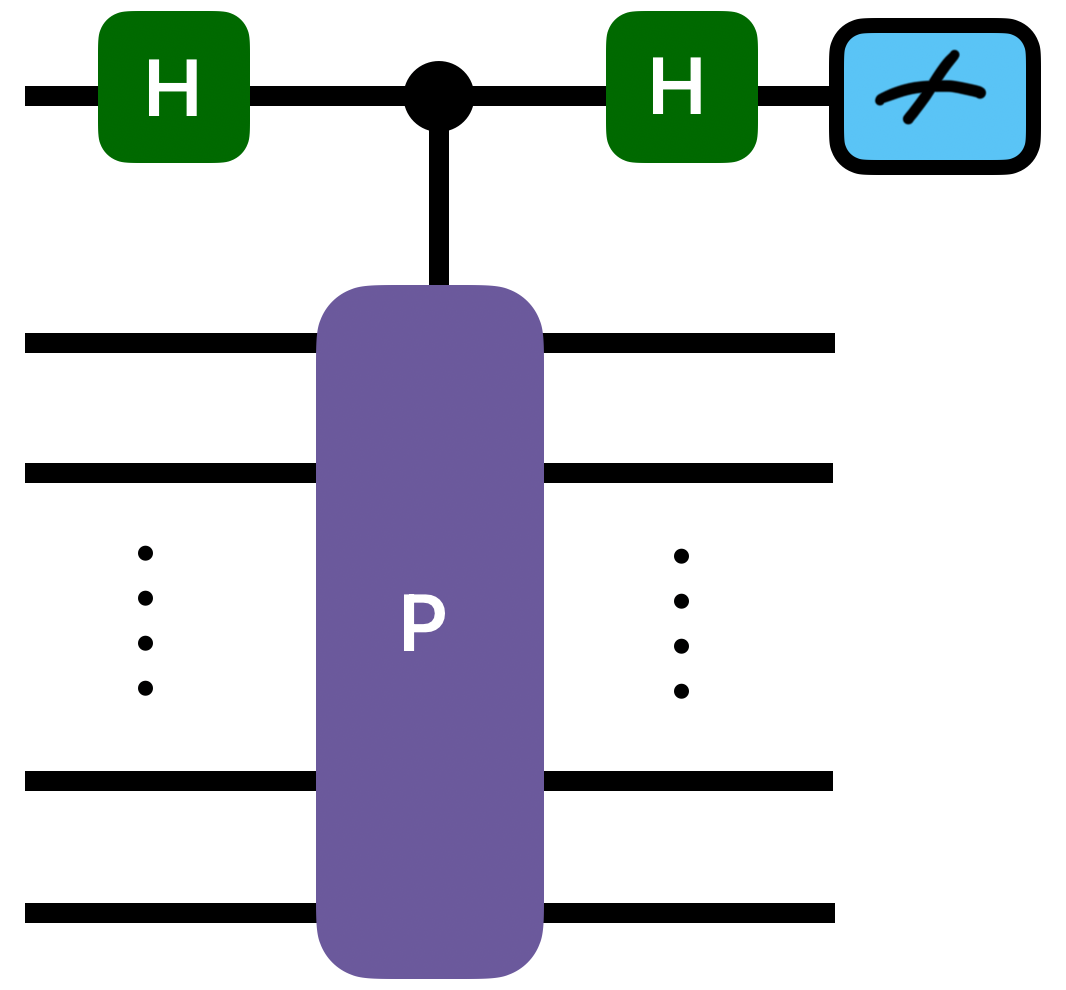}
    \caption{An illustration of the circuit that can used to prepare a stabilizer state. The purple box labeled ``P'' represents a multi-qubit Pauli gate and the green boxes labeled ``H'' represent the Hadamard gate.}
    \label{fig:Cliff_cir}
\end{figure}

Given a $n$-qubit stabilizer state $\ket{\psi_s} = (\mathcal{I}+ (-1)^l E )\ket{\psi_{s}^{'}}/\sqrt{2}$, where $E$ is a Pauli-string of only $X$s and $\psi_{s}^{'}$ is another $n$-qubit stabilizer state, we now construct an unitary, $U$, on $n+1$-qubit, such that $U\ket{\psi_{s}^{'}} = \ket{\psi_{s}}$.
If the excitation generator $E$ is of the form $X_{k_1}X_{k_2}...$, the unitary, $U$, is:
\begin{equation}\label{eq:stab_u}
    U = \text{H}(a) \bigg(\prod_{k_1, k_2 ...} \text{CNOT} (a, k_i) \bigg)\text{H}(a),
\end{equation}
where H($a$) is the Hadamard gate acting on the ancilla qubit-$a$ and CNOT$(a, k_i)$ is the Controlled-NOT gate with $a$ being the control qubit and $k_i$ being the target qubit.
A schematic depiction of $U$ is shown in Fig.~\ref{fig:Cliff_cir}.
It can be seen that $U (\ket{0}\otimes \ket{\psi_{s}^{'}}) =1/\sqrt{2} \{\ket{0} \otimes [(\mathcal{I} + E)\ket{\psi_{s}^{'}}/\sqrt{2}] + \ket{1} \otimes [(\mathcal{I} - E)\ket{\psi_{s}^{'}}/\sqrt{2}] \}$, which upon measuring the ancilla qubit results in the state, $\ket{\psi} = \ket{l}\otimes\ket{\psi_s} = \ket{l}\otimes(\mathcal{I}+ (-1)^l E )\ket{\psi_{s}^{'}}/\sqrt{2}$, where $l$ is the measurement outcome of the ancilla qubit.

We note that using unitaries of the form in Eq.~\ref{eq:stab_u} one can generate any stabilizer state of the form in Eq.~\ref{eq:iter_stab} or Eq.~\ref{eq:ad_stab}.

\subsection{Generalized stabilizer states}\label{sec:gen_stab_s}
\begin{figure}[htbp!]
    \centering
    \includegraphics[width=0.75\linewidth]{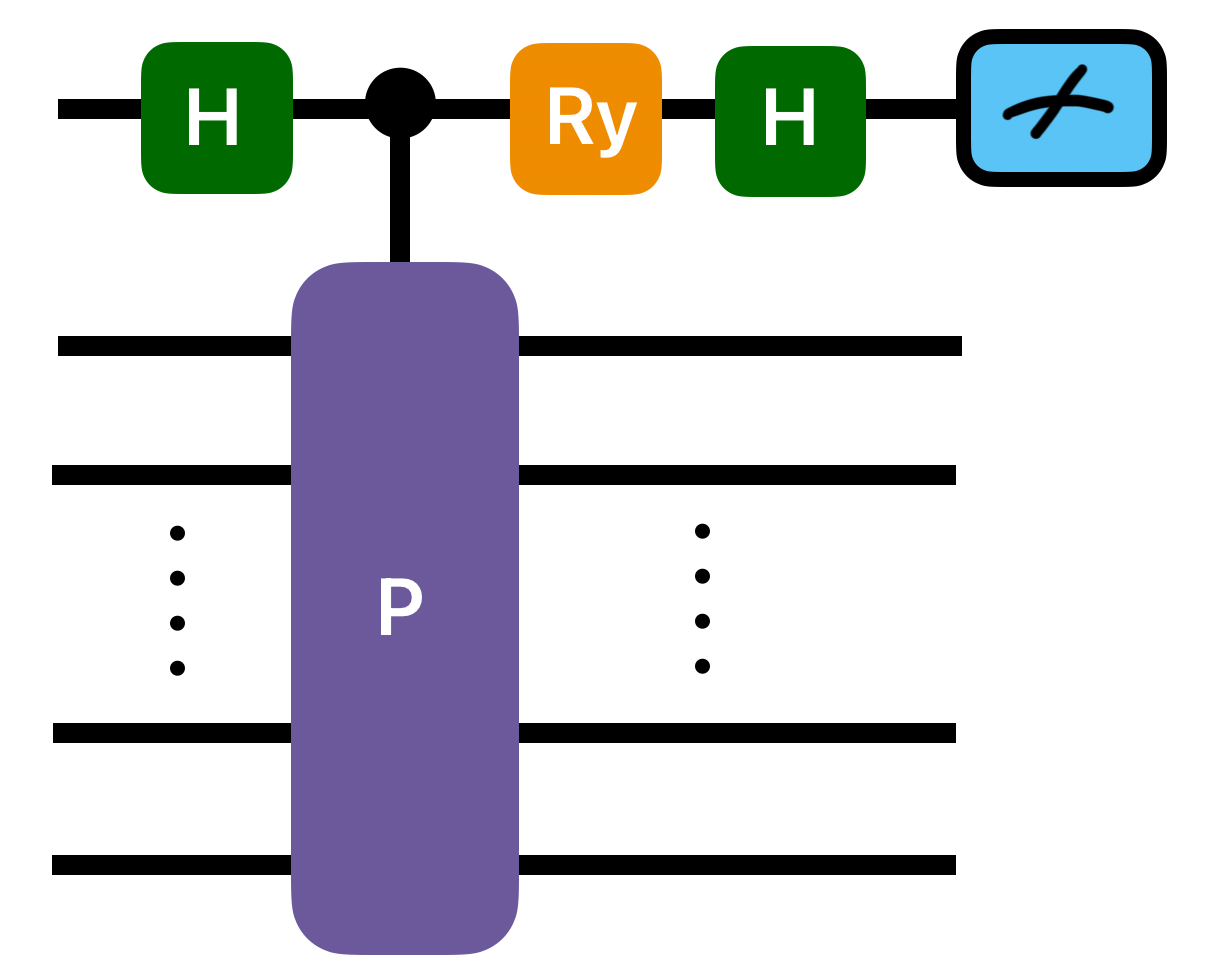}
    \caption{An illustration of the circuit that can used to prepare a generalized stabilizer state. The purple box labeled ``P'' represents a multi-qubit Pauli gate, the green boxes labeled ``H'' represent the Hadamard gate and the orange box labeled ``Ry'' represents a Ry($\theta$) gate (rotation around the Y-axis by an angle $\theta$).}
    \label{fig:gen_Cliff_cir}
\end{figure}

We can further modify the best stabilizer states, $(\mathcal{I} + (-1)^l E)/\sqrt{2}\ket{\psi_{s}^{'}}$, by allowing for arbitrary amplitudes when forming the superposition as follows:
\begin{equation}\label{eq:gen_stab}
    \ket{\psi_{s}} = (x\mathcal{I} + y(-1)^l E)\ket{\psi_{s}^{'}},
\end{equation}
where $x$ and $y$ are real amplitudes and $x^2+y^2=1$.
We do this by adding an extra gate, a single qubit rotation gate around the Y-axis, $\text{Ry}(\theta)$ (depicted by the orange box in Fig.~\ref{fig:gen_Cliff_cir}) on the ancilla qubit.
The parameter, $\theta$, can be optimized to minimize the energy corresponding to the state.

\begin{figure*}[htbp!]
    \centering
    \begin{tabular}{c c c}
    \toprule
    a) H$_{2}$ molecule   & b)  H$_4$ molecule & c)  LiH molecule\\
    \midrule
    \multicolumn{3}{c}{\textbf{Energy}}\\
    \midrule
    \includegraphics[width=0.67\columnwidth]{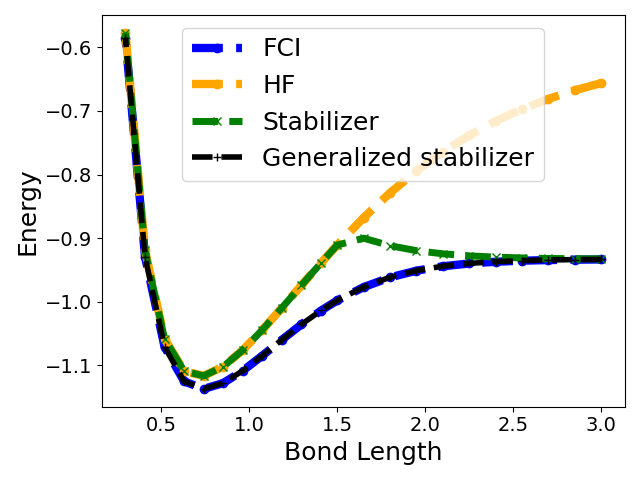} & 
    \includegraphics[width=0.67\columnwidth]{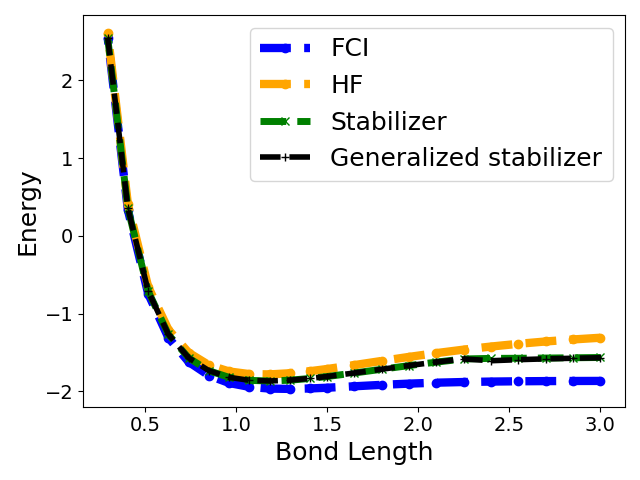} &
    \includegraphics[width=0.66\columnwidth]{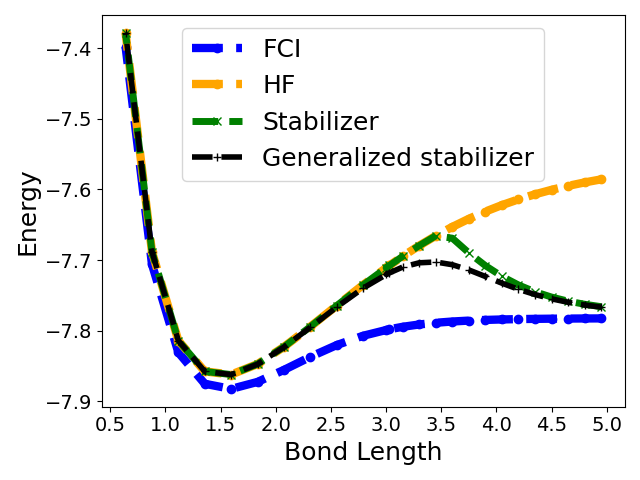}\\
    \midrule
    \multicolumn{3}{c}{\textbf{Error in energy}}\\
    \midrule
    \includegraphics[width=0.67\columnwidth]{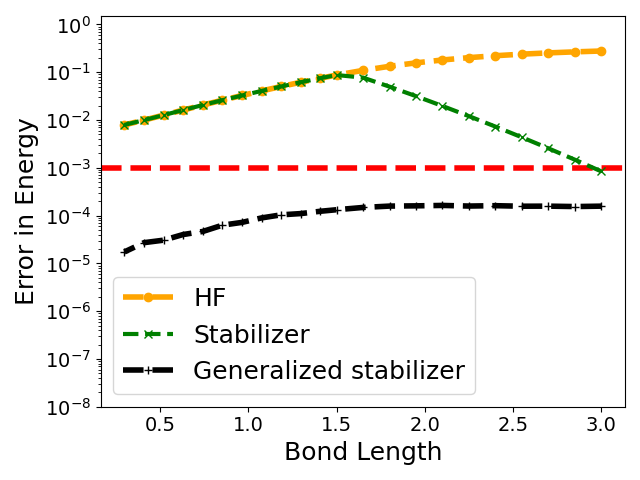} & 
    \includegraphics[width=0.67\columnwidth]{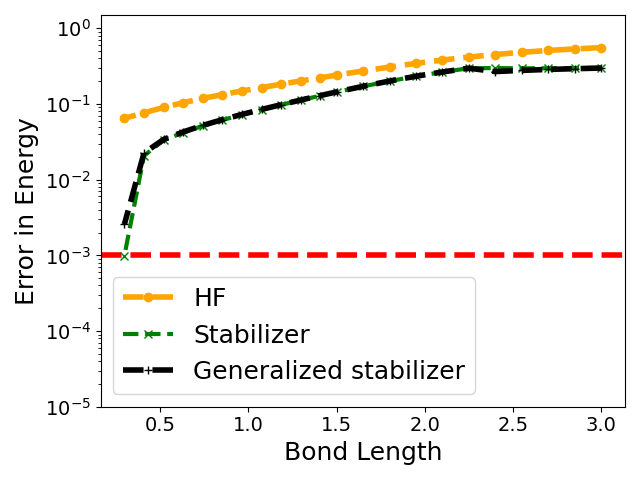} & 
    \includegraphics[width=0.66\columnwidth]{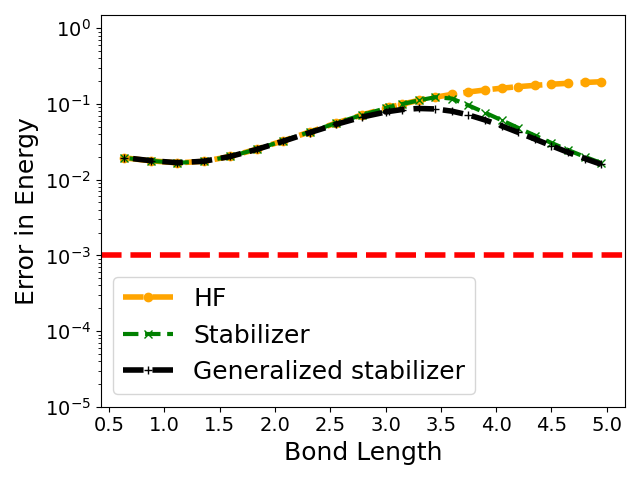}\\
    \midrule
    \toprule
    d) BH$_{3}$ molecule   & e)  N$_2$ molecule & f)  BeH$_2$ molecule\\
    \midrule
    \multicolumn{3}{c}{\textbf{Energy}}\\
    \midrule
    \includegraphics[width=0.67\columnwidth]{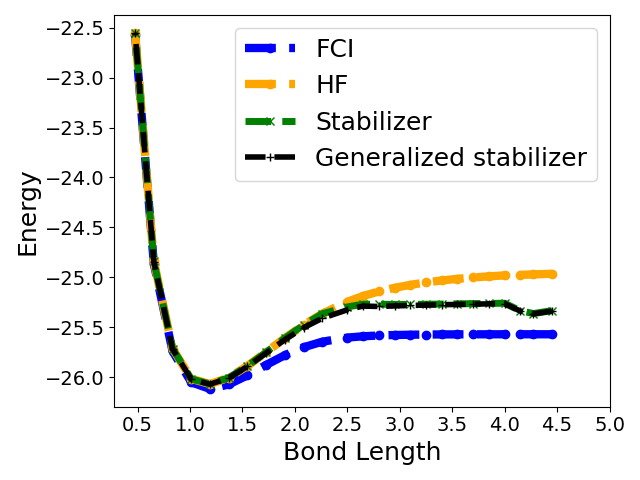} & 
    \includegraphics[width=0.67\columnwidth]{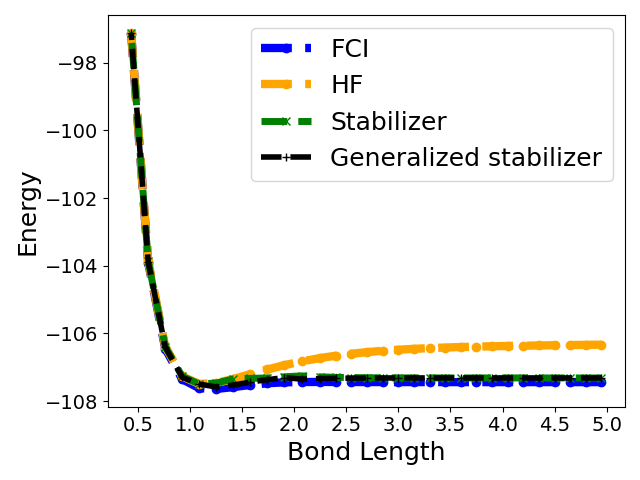} &
    \includegraphics[width=0.66\columnwidth]{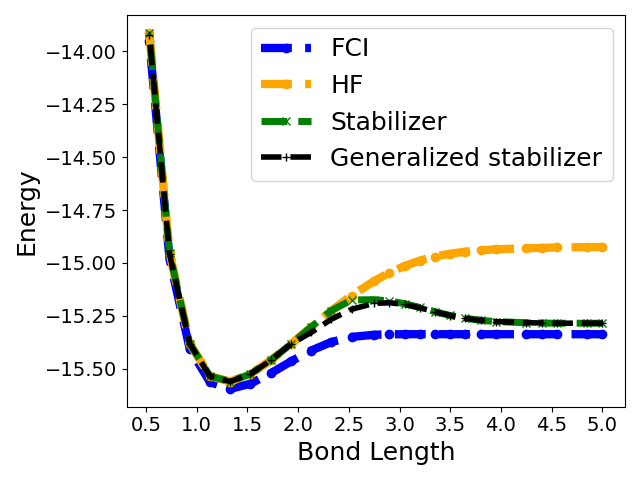}\\
    \midrule
    \multicolumn{3}{c}{\textbf{Error in energy}}\\
    \midrule
    \includegraphics[width=0.67\columnwidth]{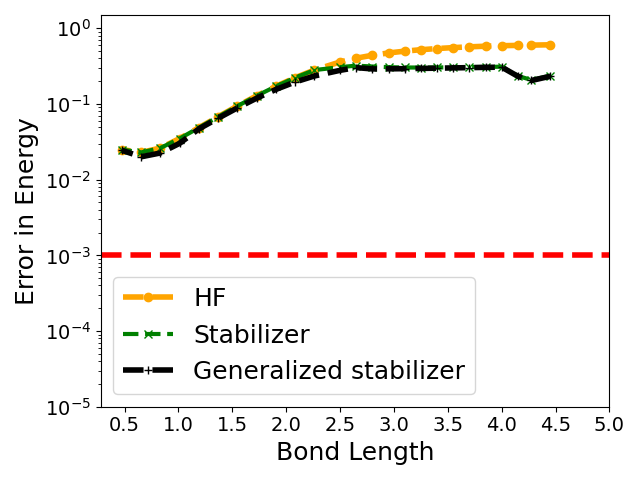} & 
    \includegraphics[width=0.67\columnwidth]{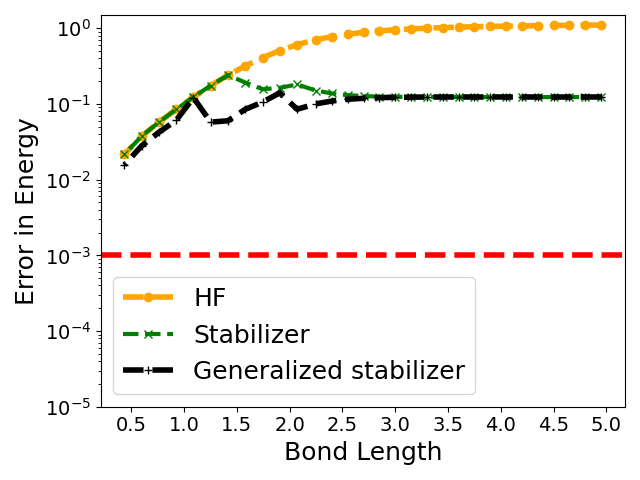} & 
    \includegraphics[width=0.66\columnwidth]{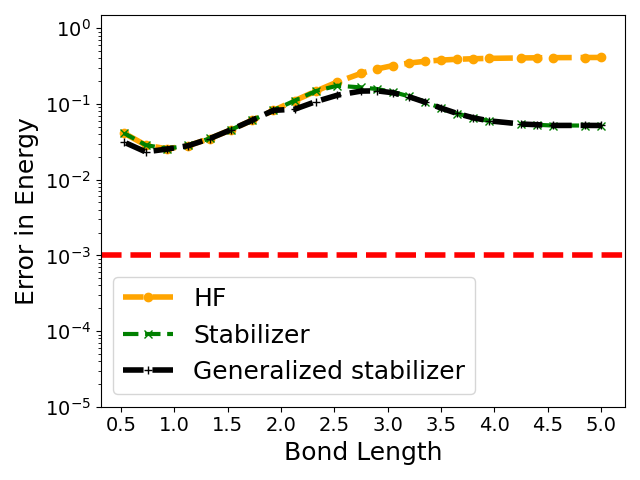}\\
    \midrule
    \end{tabular}
    \caption{\label{fig:med_mol} 
    A plot showing the potential energy surface (top) and error when compared to the FCI energy (bottom) for different molecules considered in this study. The red line denotes an error value of 1mHa.} 
\end{figure*}

We use full stabilizer CI algorithm (Sec.~\ref{sec:stab_CI}) for finding the stabilizer approximation to the ground state of these molecules.
We also find the generalized stabilizer states corresponding to the best stabilizer states by following the procedure in Sec.~\ref{sec:gen_stab_s}.
The results from the different numerical experiments are presented in Fig.~\ref{fig:med_mol} and Fig.~\ref{fig:big_mol}.

\subsection{Results}
In this section, we present the result from various numerical simulations carried to find stabilizer approximation for different molecules.
We use the python packages tequila~\cite{kottmann2021tequila}, stim~\cite{gidney2021stim} and cirq~\cite{cirq_developers_2024_11398048} to perform the various calculations.

First we use the stabilizer CI (Sec. \ref{sec:stab_CI}) method for small molecules, followed by the adaptive stabilizer method (Sec. \ref{sec:ad_stab_CI}) for larger molecules and report the results next.
All energy values are in Hartree (Ha) units and all bond length values are in Angstrom (\AA) units,  unless specified otherwise.
All output files, source codes, and example simulations are available at~\cite{github_StabCI}.

All the simulations were performed on a single node of the Duke Compute Cluster. 
The results presented in the sections below required compute times ranging from a few seconds for the smallest molecule (H$_2$) to approximately five hours for the largest molecules (Cr$_2$), representing a significant speedup over previous methods~\cite{ravi2022cafqa}.

\subsubsection{Small molecules}\label{sec:sml_m}
We consider different commonly studied small molecules in the minimal basis (STO-3G), such as H$_2$, LiH, H$_4$, BH$_3$, BeH$_2$ and N$_2$.
The details of the different molecules are summarized in Table~\ref{tab:sm_mol_det}.

\begin{table}[htbp!]
    \centering
    \begin{tabular}{c | c | c}
        \hline
        Molecule & Number of Electrons & Number of Qubits \\
        \hline
        \hline
        H$_2$ & 2(2) & 4(4)\\ 
        \hline
        LiH & 4(4) & 12(12) \\ 
        \hline
        H$_4$ & 4(4)  & 8(8) \\ 
        \hline
        N$_2$ & 6(14) &  12(20) \\ 
        \hline
        BeH$_2$ & 6(6)  & 14(14) \\ 
        \hline
        BH$_3$ & 6(8) & 12(16) \\ 
        \hline
        \hline
    \end{tabular}
    \caption{A table containing the details of the different molecules considered here. The numbers correspond to the used (total) number of electrons/qubits, respectively.}
    \label{tab:sm_mol_det}
\end{table}

It can be seen from Fig.~\ref{fig:med_mol}, that the HF state is a pretty good approximation to the actual ground state in configurations close to the equilibrium geometry of the molecules.
However, they tend to perform worse for stretched configurations far from equilibrium, which is the expected, because of the lack of electronic correlations in these states.

The stabilizer state with the lowest energy in configurations close to the equilibrium geometry of the molecules is the HF state~\cite{anand2023hamiltonians, anand2023leveraging} and the proposed stabilizer CI method also identifies this state as the optimal stabilizer approximation.
However, for the case of the Hydrogen ring (H$_4$ molecule - Fig.~\ref{fig:med_mol} b), we observe that the stabilizer state with the lowest energy is not the HF state, but an entangled state.

In stretched configurations far from equilibrium, the proposed method finds entangled stabilizer states as the best stabilizer approximation.
These states capture some of the electronic correlations and thus have energies closer to the true ground state.
We see close to an order magnitude reduction in error when compared to the HF state, for all molecules in highly stretched configurations.
In certain cases (H$_4$ and H$_2$ molecule), the energy corresponding to these states are well within the chemical accuracy (error < 1 mHa).

Finally, it should be noted that the generalized stabilizer state is better approximation to the ground state than the stabilizer state, as it allows for a biased superposition of some excitations.
In case of H$_2$ molecule, (Fig.~\ref{fig:med_mol} a) the generalized stabilizer state is a very good approximation (error < 1 mHa) to the true ground state for all configurations.
This high accuracy is due to the fact that both the true ground state and the optimal stabilizer state share the same configuration. 
By removing the unbiased restriction inherent in stabilizer states, we can optimize the amplitudes, thereby bringing the approximation closer to the true ground state.

This is not the case for all other molecules and the generalized stabilizer state only leads to small improvements in energy.
Nonetheless, this provides evidence that stabilizer states with similar configuration to the ground state can be a good starting point for finding ground states.

\subsubsection{Larger molecules}\label{sec:lar_m}
\begin{figure}[htbp!]
    \centering
    \begin{tabular}{c}
    \toprule
    a) C$_2$H$_6$ molecule \\
    \midrule
    \includegraphics[width=0.99\columnwidth]{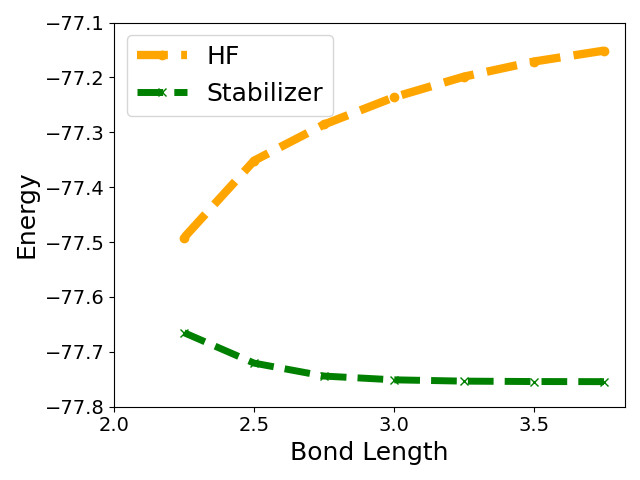} \\
    \midrule
    b)  Cr$_2$ molecule \\
    \midrule
    \includegraphics[width=0.99\columnwidth]{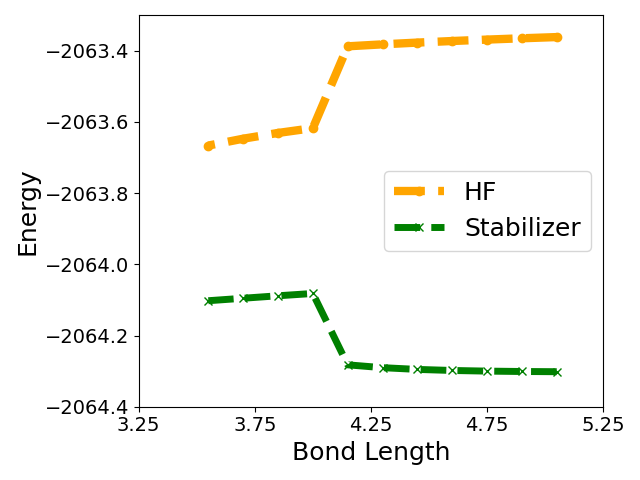}\\
    \midrule
    \end{tabular}
    \caption{\label{fig:big_mol} A plot showing the energy values of the Hartree-Fock state and best stabilizer state for different bond length of the C$_2$H$_6$ and Cr$_2$ molecule.} 
\end{figure}

We now report the result for two slightly larger molecules C$_2$H$_6$ and Cr$_2$ in the minimal (STO-3G) basis.
We consider an active space of 28 qubits (14 occupied (electrons) and 14 unoccupied spin orbitals) for the C$_2$H$_6$ and 36 qubits (12 occupied (electrons) and 24 unoccupied spin orbitals) for the Cr$_2$ molecule.
These molecules are significantly larger than the ones considered in Sec.~\ref{sec:sml_m}, thus we do not use the full stabilizer CI algorithm (Sec.~\ref{sec:stab_CI}) but the adaptive stabilizer CI algorithm (Sec.~\ref{sec:ad_stab_CI}) for finding the stabilizer approximation to the ground state of these molecules.
We plot the result from the numerical experiments in Fig.~\ref{fig:big_mol}.

We plot the results for molecular geometries where the bond length exceeds twice the equilibrium geometry, as it is well-known that the Hartree-Fock (HF) state performs poorly in these regions.
As shown in Fig~\ref{fig:big_mol}, the best stabilizer states (see the Appendix~\ref{app:ex_code} for some examples) consistently achieve lower energies compared to the HF state. 
This behavior is similar to our observations for smaller molecules, where introducing excitations to the HF state improves the approximation as they capture electronic correlations.

These results suggest that stabilizer states can serve as reliable approximations to the true ground states of molecules, providing a better starting point for more complex simulations, such as VQE~\cite{peruzzo2014variational} or QPE~\cite{aspuru2005simulated, abrams1997simulation}, in situations where the HF state is not a good approximation.

\section{Error Detection}\label{sec:codes}
In this section, we provide a detailed description of the procedure for constructing the codes and demonstrate their error-detection capabilities.

We first note that any state $\ket{\psi}$ that we construct as an approximation of the ground state in this study (using Eq.~\ref{eq:iter_stab} or Eq.~\ref{eq:ad_stab}) is a stabilizer state.
Thus, there exist a set of operators $\mathcal{S}$, such that $s_i \ket{\psi} = \ket{\psi} \forall s_i \in \mathcal{S}$.
The exact  procedure to construct the stabilizers is presented in detail in the Appendix (\ref{app:stb_ls}).

Next, we outline the procedure of constructing a error detection code which can be used to prepare the state $\ket{\psi}$.
\begin{enumerate}
    \item Given the stabilizers $\mathcal{S}$ of the state $\ket{\psi}$, find the local Clifford unitary, $\mathcal{C}$, that converts it to the standard form, $\mathcal{S}^{'}$.~\cite{Van_den_Nest_2004, zeng2007local, anand2023hamiltonians} (see Sec.~\ref{sec:stab_state} for details)
    \item Construct the error characterization table for this stabilizer group and find the (abelian) operators that can be used as word operators, $\mathcal{W}$. (see Appendix~\ref{app:err_wor} for details)
    \item Define a Codeword stabilized (CWS) code using the word operators, $\mathcal{W}$, and the stabilizers, $\mathcal{S}^{'}$.
    \item Find the equivalent stabilizer code as follows:
    \begin{enumerate}
        \item Fix one of the non-trivial word operator, $w_i$, to be the logical-$X$ operator.
        \item Find a stabilizer $s_i \in \mathcal{S}^{'}$ that anti-commutes with, $w_i$, and fix it to be the logical-$Z$ operator.
        \item  Remove $s_i$ from $\mathcal{S}^{'}$ and modify the remaining stabilizers to get a set of stabilizers $\mathcal{S}^{''}$, so that every remaining stabilizer commutes with $w_i$ and $s_i$.
        \item The stabilizers $\mathcal{S}^{''}$, logical-$X = w_i$ and logical-$Z = s_i$ operators, define the equivalent stabilizer code.
    \end{enumerate}
\end{enumerate}

We use the above procedure to construct some error detection codes for different molecules studied in paper and present them in the Appendix (\ref{app:ex_code}).
The resulting code can be used to prepare the best stabilizer state $\ket{\psi}$ of the form in  Eq.~\ref{eq:iter_stab} or Eq.~\ref{eq:ad_stab} with error detection on a physical device and post-selection, as the state $\ket{\psi}$ is a logical basis state of the code.

In certain cases (usually states with more than 2 excitations), we observe that some generalized stabilizer state (of the form in Eq.~\ref{eq:gen_stab}) can also be prepared while detecting errors and using post-selection.
This is possible because in these cases, the generalized state are the arbitrary superposition of the logical states, ($\ket{\Bar{0}}$ and $\ket{\Bar{1}}$) or ($\ket{\Bar{+}}$ and $\ket{\Bar{-}}$).
Thus, using protocols that allow for universal quantum computations, such as magic state distillation~\cite{bravyi2005universal}, we can also prepare these states with the codes.
Some examples of such states are presented in the Appendix (\ref{app:ex_gen_code}).

This suggests that if the code we construct can have more logical states with the same symmetry as that of the molecule, we can prepare more generalized stabilizer states. 

Next, we carry out numerical simulations with noise to demonstrate the preparation of a (generalized) stabilizer state for the case of H$_4$ molecule using the codes we construct, which are listed in the Appendix~\ref{app:ex_code}.
We prepare a stabilizer state, $\ket{\psi_s}$,
\begin{align}
    \ket{\psi_s} &= \frac{1}{2}(\ket{11110000} - \ket{11000011} \nonumber\\ 
    &- \ket{00111100} + \ket{00001111}), \nonumber
\end{align}
and an arbitrary generalized stabilizer state, $\ket{\psi_s^{'}}$,
\begin{align}
    \ket{\psi_s^{'}} &= 0.3928\ket{11110000} - 0.5879\ket{11000011} \nonumber\\
    &- 0.5879\ket{00111100} + 0.3928\ket{00001111}. \nonumber
\end{align}

\begin{figure}[htbp!]
    \centering
    \begin{tabular}{c}
    a) Noisy circuit \\
    \includegraphics[width=0.75\columnwidth]{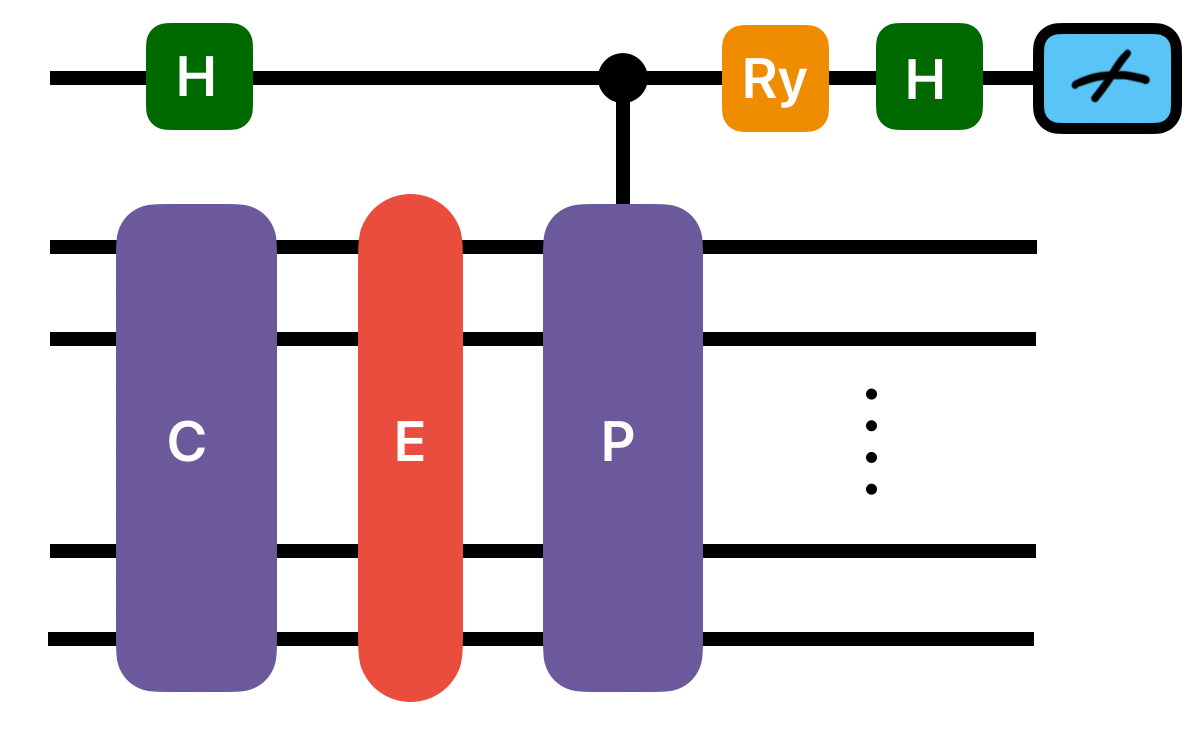} \\
    b)  Noisy circuit with stabilizer measurements \\
    \includegraphics[width=0.99\columnwidth]{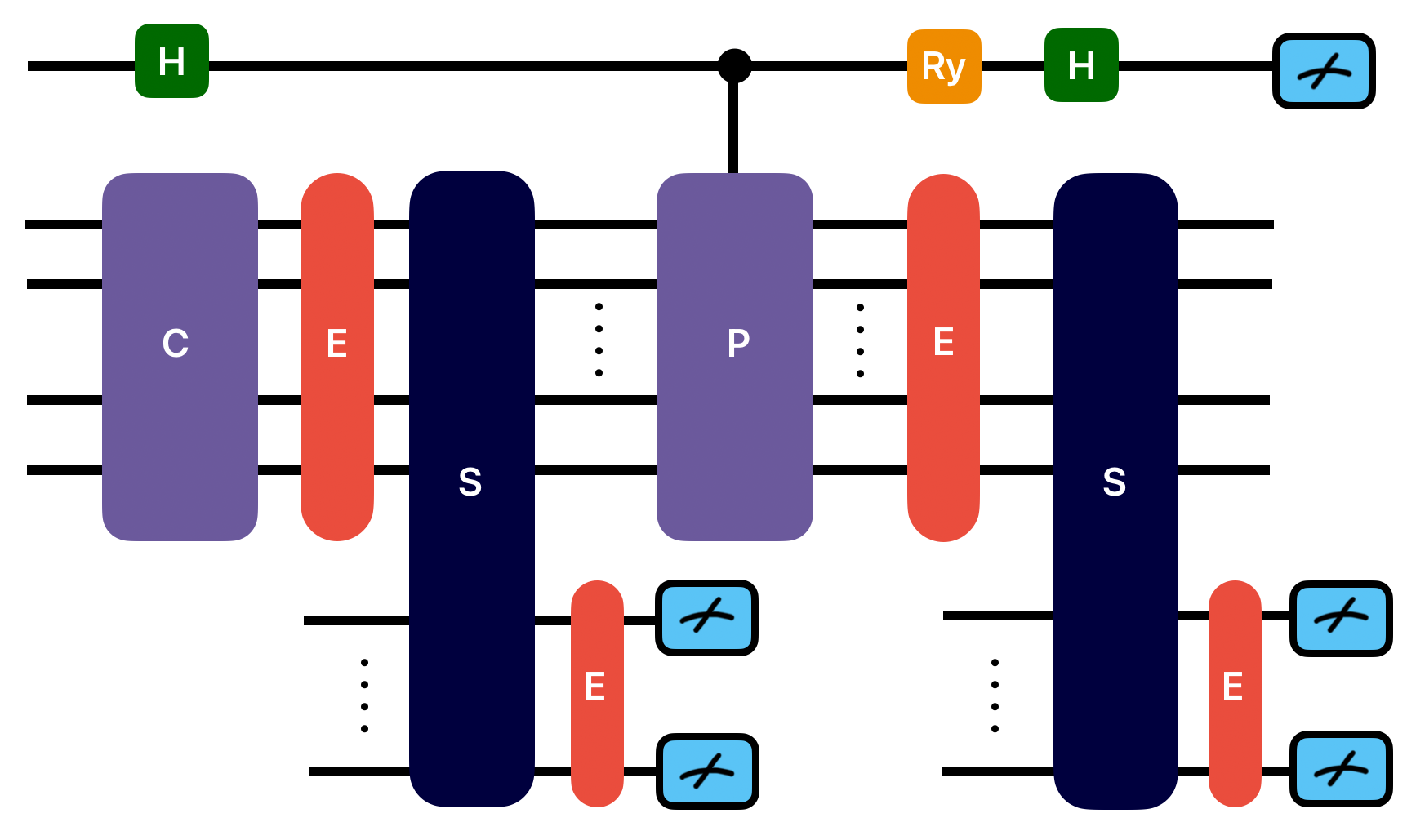}\\
    \end{tabular}
    \caption{\label{fig:noisy_sim} Illustration of circuits used for noisy preparation of a (generalized) stabilizer state in the case of the H$_4$ molecule. The purple boxes labeled ``C'' and ``P'' represent a Clifford circuit and multi-qubit Pauli gate, respectively. The green boxes labeled ``H'' represent the Hadamard gate, the orange box labeled ``Ry'' represents a Ry($\theta$) gate (rotation around the Y-axis by angle $\theta$), the dark blue boxes labeled ``S'' represent the stabilizer extraction circuit and the red boxes labeled ``E'' represent the error (depolarizing or bit-flip) channel.} 
\end{figure}

To perform noisy simulation, we use a simple noise model, where we add a depolarizing channel after state preparation circuit and bit-flip channel before stabilizer measurement.
An illustration of the circuits used in these simulation is shown in Fig.~\ref{fig:noisy_sim}.
The gates marked in red color represent the depolarizing channel on the data qubits and bit-flip channel on the ancilla qubits.
In all the simulations, we fixed bit-flip probability to be half the depolarizing probability.
All other operations in these circuits are ideal.
The actual circuit is shown in the Appendix~\ref{app:noisy_cir}.

\begin{figure}[htbp!]
    \centering
    \begin{tabular}{c}
    a) Stabilizer state \\
    \includegraphics[width=0.95\columnwidth]{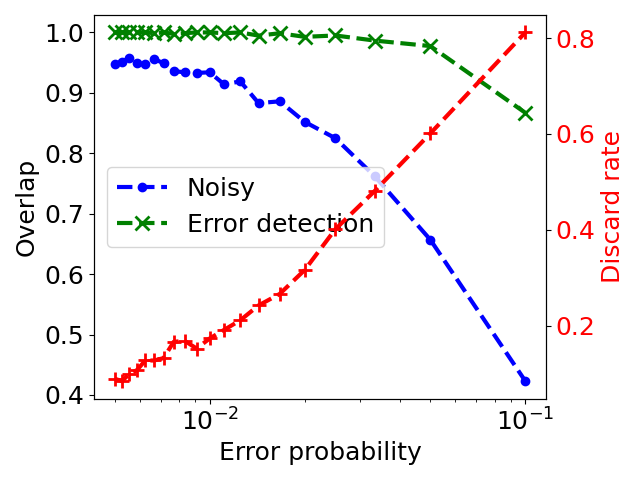} \\
    b)  Generalized stabilizer state \\
    \includegraphics[width=0.95\columnwidth]{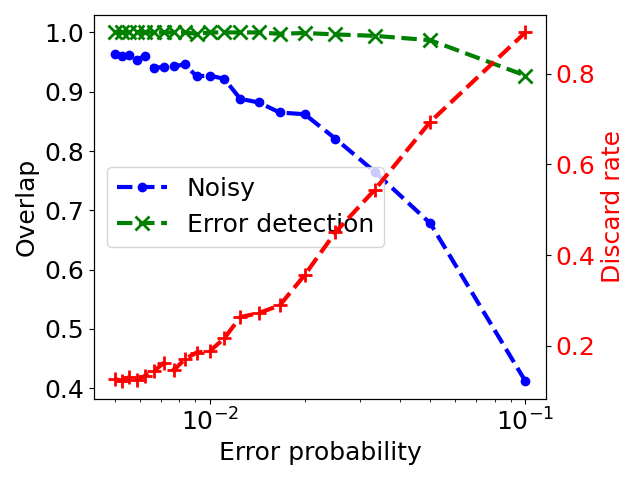}\\
    \end{tabular}
    \caption{A plot comparing the overlap of noisy states with the ideal state from simulations with and without error detection as a function of the depolarizing error probability. The red line represents the discard rate for the simulations with error detection.}
    \label{fig:noisy_states}
\end{figure}

Using the circuits in Fig.~\ref{fig:noisy_sim}, we prepare noisy states and calculate the overlap with the ideal states for different error probabilities.
In the noisy simulations, with and without error detection, we use 1000 different repetitions of the circuits to calculate the average overlap.
In the noisy simulations with stabilizer measurements, we discard any state if we detect any error.
The results from these simulations are plotted in Fig.~\ref{fig:noisy_states}.

In simulations without error detection, we observe that the overlap decreases exponentially as a function of the error probability. 
Using error detection, we achieve a higher overlap compared to unprotected noisy operations, albeit at the cost of an increasing discard rate as the error probability rises. 
However, it should be noted that for error probabilities below 0.01, the overlap remains close to 1, with a discard rate of less than 20\%. 
Therefore, our proposal can be particularly useful in these regimes.

\section{Conclusion}\label{sec:conclusion}
We have proposed a novel framework for finding molecular subspaces with error detection properties by integrating the design of algorithms and error detection codes.
By focusing on ground state approximation, we developed efficient classical algorithms for constructing stabilizer approximations to the ground states of various molecules.
Additionally, we introduced circuit constructions for preparing generalized stabilizer states.

We then use the stabilizer approximations to construct quantum error detection codes using codeword stabilized codes. 
These codes require minimal resources and facilitate state preparation with error detection for both stabilizer and some generalized stabilizer states.

We also conduct various numerical experiments to verify the proposed method by investigating molecules up to 36 qubits. 
We observe that the stabilizer states outperform the Hartree-Fock state in every case. 
However, while they do not yet achieve the desired accuracy for most molecules studied, we find that these stabilizer states can serve as a promising starting point for further complex calculations.

Using noisy numerical simulations, we verify the error detection properties of the codes we construct for preparing both a stabilizer and a generalized stabilizer state.
These simulations suggest that our method can be useful in preparing approximate eigenstates for different molecular systems, albeit with a measurement overhead due to increasing discard rates.

This work represents an initial step toward algorithms suitable for implementation on early fault-tolerant quantum devices, with a focus on single stabilizer states.
We anticipate that employing linear combinations~\cite{childs2012hamiltonian} of different stabilizer states could yield better ground state approximations, similar to studies using superposition of spin-coupled states~\cite{marti2024spin} for accurate ground state energies. 
Further research is needed to explore this approach, develop fault-tolerant protocols for preparing these states, and extend them to universal states using recent protocols~\cite{delfosse2025low, khitrin2025unbiased}.
Future work will also examine the error-correction properties of different physical Hamiltonians.

\section*{Acknowledgements}
The authors thank Andrew Nemec and Bálint Pató for helpful discussions.
This work was supported by the National Science Foundation (NSF) Quantum Leap Challenge Institute of Robust Quantum Simulation (QLCI grant OMA-2120757).

\bibliography{main.bib}

\clearpage
\newpage
\onecolumngrid
\section*{Appendix}\label{sec:Appendix}

\subsection{Stabilizers of unbiased sum of stabilizer states}\label{app:stb_ls}
The procedure outlined in this section closely follows Ref.~\cite{garcia2017geometry}.

Given any stabilizer state, $\ket{\psi_{i}}$, (reference state in Eq.~\ref{eq:iter_stab} or Eq.~\ref{eq:ad_stab}), 
there exist a Clifford unitary, $\mathcal{C}_i$, which maps the state, $\ket{\psi_{i}}$, to basis from, i.e., $\mathcal{C}_i\ket{\psi_{i}} = \ket{0}^{\otimes n}$.

It can also be verified that the action of $\mathcal{C}_i$ on $E_{i+1}\ket{\psi_{i}}$ also produces a basis state, 
$\mathcal{C}_iE_{i+1}\ket{\psi_{i}} = \mathcal{C}_iE_{i+1}\mathcal{C}_{i}^{\dagger}\mathcal{C}_i\ket{\psi_{i}} = \mathcal{C}_iE_{i+1}\mathcal{C}_{i}^{\dagger}\ket{0}^{\otimes n} = E_{i+1}^{'}\ket{0}^{\otimes n} = i^m \ket{b_1 b_2 ... b_n}$, where $m \in \{0,1,2,3\}$, $b_i \in \{0,1\}$ and $E_i, E_{i+1}^{'} \in \mathcal{G}_n$.
Thus, the state $\ket{\psi_{i+1}}$ in Eq.~\ref{eq:iter_stab} or Eq.~\ref{eq:ad_stab} can be written as:

\begin{align}
    \ket{\psi_{i+1}} &=  \mathcal{C}_i^\dagger\bigg(\frac{\ket{0}^{\otimes n} + i^{(2l+m) \text{mod} 4} \ket{b_1 b_2 ... b_n}}{\sqrt{2}}\bigg) \nonumber\\
    &=  \mathcal{C}_i^\dagger \ket{\psi_z} \ket{\psi_{\text{GHZ}}},
\end{align}

where $\ket{\psi_z} $ is an all-zero state supported on qubits, $ i \in \{{i|b_i = 0}\}$ and $\ket{\psi_{\text{GHZ}}}$ is the GHZ state supported on qubits, $ i \in \{{i|b_i = 1}\}$.

The stabilizers of the state $\ket{\psi_{i+1}}$ can then be written as $\mathcal{S} =  \mathcal{C}_i^\dagger \langle \mathcal{S}_z, \mathcal{S}_{\text{GHZ}} \rangle  \mathcal{C}_i$, where $\mathcal{S}_z = \langle Z_i, i \in \{{i|b_i = 0}\}\rangle$ and $\mathcal{S}_{\text{GHZ}}$ is supported on qubits, $ i \in \{{i|b_i = 1}\}$ and 
\begin{align}
    \mathcal{S}_{\text{GHZ}} &= \begin{cases}
        \langle (-1)^{t/2} XX...X , \forall i Z_i Z_{i+1} \rangle &\text{if $t$ = 0 }\\
        \langle (-1)^{(t-1)/2} YY...Y , \forall i Z_i Z_{i+1}\rangle  &\text{if $t$ = 1} \nonumber\\
    \end{cases} 
\end{align}
where $t = ((2l+m)$ mod 4) mod 2.

Also, the Clifford unitary $\mathcal{C}_{i+1}$ that maps $\ket{\psi_{i+1}}$ to basis from, $\ket{0}^{\otimes n}$ can be written as:
\begin{equation}
    \mathcal{C}_{i+1} = \mathcal{C}_{\text{GHZ}} \mathcal{C}_i,
\end{equation}
where $\mathcal{C}_{\text{GHZ}}$ is the Clifford unitary that maps the GHZ state $\ket{\psi_{\text{GHZ}}}$ to the state $\ket{0}^{\otimes m}$.

\subsection{Error characterization table and word operators}\label{app:err_wor}
Given a set of stabilizers, $\mathcal{S}$, in the standard form, i.e., every element $s_i$ is of the form $X_{u_{i}}Z_{\{v_{i}\}}$, where, $u_{i}$ is an integer and $\{v_{i}\}$ is a set of integers.
It can be noted that any single qubit error is equivalent to an error-string of only $Z$s.
This can be seen by representing them in a tabular form, which we refer to as the error-characterization table (see Table~\ref{tab:err_char}).

\begin{table}[htbp!]
    \centering
    \begin{tabular}{c|c|c|c|c}
     Error$\backslash$ Qubit & 1 & 2 & ... & n \\
     \midrule
       $X_{i}$  & $Z_{\{v_1\}}$ & $Z_{\{v_2\}}$ & ... & $Z_{\{v_n\}}$ \\       
       $Y_{i}$  & $Z_1$ $Z_{\{v_1\}}$ & $Z_2$ $Z_{\{v_2\}}$ & ... & $Z_{n}Z_{\{v_n\}}$ \\ 
       $Z_{i}$  & $Z_1$ & $Z_2$ & ... & $Z_n$ \\
    \end{tabular}
    \label{tab:err_char}
    \caption{An illustration of the error characterization table.}
\end{table}

There are 2$^n$-1 distinct Pauli operators of only $Z$s and at-most 3$n$ different single qubit errors. 
Thus, for $n > 3$, there exists an operator, $w$, that is not present in the error characterization table.
The set of operators, $\{\mathcal{I}, w\}$, along with the stabilizers, $\mathcal{S}$, can be used to construct a codeword stabilizer code.

\subsection{Example stabilizer states and corresponding codes}\label{app:ex_code}
In this section, we present some example quantum error detection codes for preparing stabilizer ground state approximations of different molecules that were considered in this study.

\subsubsection{H$_2$ molecule}
The stabilizer state, $\ket{\psi_s}$, with the lowest energy at bond length 3.0\AA \hspace{0.1mm} is,\begin{align}
  \ket{\psi_s} &= \frac{1}{\sqrt{2}}(\ket{1100} -\ket{0011}).\nonumber
\end{align}
The stabilizers for this state are
\begin{align}
    \mathcal{S} = \{ -Z_1Z_3, -Z_2Z_4, Z_3Z_4, -X_1X_2X_3X_4\}. \nonumber
\end{align}
Using the operators $\{ \mathcal{I}, Z_3Z_4\}$ as word operators, the resulting stabilizer code is a [[4,1,2]] code and has the following logical operators, $\Bar{Z} = Z_3Z_4$, $\Bar{X} = -X_1X_3$, and the following stabilizers, $\mathcal{S}^{'} = \{ -Z_1Z_3, -Z_2Z_4, -X_1X_2X_3X_4 \}$.
It can be verified that the state, $\ket{\psi_s}$, is in the codespace of the above stabilizer code.

\subsubsection{H$_4$ molecule}
The stabilizer state, $\ket{\psi_s}$, with the lowest energy at bond length 3.0\AA \hspace{0.1mm} is,
\begin{align}
    \ket{\psi_s} &= \frac{1}{2}(\ket{11110000} - \ket{11000011} - \ket{00111100} + \ket{00001111}). \nonumber
\end{align}

The stabilizer code for the above state is a [[8, 1, 2]] code and has the following logical operators, $\Bar{X} = Z_6Z_8$, $\Bar{Z} = -X_1X_2X_5X_6$, and the following stabilizers, $\mathcal{S}^{'} = \{-Z_1Z_6,$ $-Z_2Z_6,$ $-Z_3Z_8, $ $-Z_4Z_8,$  $Z_5Z_6,$  $Z_7Z_8,$  $X_1X_2X_3X_4X_5X_6X_7X_8\}$ .

\subsubsection{BH$_3$ molecule}
The stabilizer state, $\ket{\psi_s}$, with the lowest energy at bond length 4.45\AA \hspace{0.1mm} is,
\begin{align}
    \ket{\psi_s} &= \frac{1}{\sqrt{8}} (\ket{111111000000} +\ket{110000111100} + \ket{110100111000} -\ket{111000110100}  \nonumber \\
    &-\ket{110011001100} -\ket{111100110000} -\ket{110111001000}  +\ket{111011000100}). \nonumber
\end{align}

The stabilizer code for the above state is a [[12, 1, 2]] code and has the following logical operators, $\Bar{X} = Z_8Z_9 $, $\Bar{Z} = -X_5X_6X_7X_8$, and the following stabilizers, $\mathcal{S}^{'} = \{ -Z_1, -Z_2, -Z_3Z_9,$ $-Z_4Z_{10}, $ \hspace{0.05mm} $-Z_5Z_8,$ \hspace{0.05mm} $ -Z_6Z_8, $ \hspace{0.05mm} $Z_7Z_8, $ \hspace{0.05mm} $ X_3X_9,$ $X_4X_{10}, Z_{11}, Z_{12} \}$.

\subsubsection{BeH$_2$ molecule}
The stabilizer state, $\ket{\psi_s}$, with the lowest energy at bond length 3.0\AA \hspace{0.1mm} is,
\begin{align}
  \ket{\psi_s} &= \frac{1}{\sqrt{2}}(\ket{11111100000000} -\ket{11110011000000}).\nonumber
\end{align}

The stabilizer code for the above state is a [[14, 1, 2]] code and has the following logical operators, $\Bar{Z} = Z_7Z_8$, $\Bar{X} = -X_5X_7$, and the following stabilizers, $\mathcal{S}^{'} = \{ -Z_1, -Z_2, -Z_3, -Z_4, -Z_5Z_7, -Z_6Z_8, -X_5X_6X_7X_8,$ $ Z_9, Z_{10}, Z_{11}, Z_{12}, Z_{13}, Z_{14}\}$.

\subsubsection{C$_2$H$_6$ molecule}
The stabilizer state, $\ket{\psi_s}$, with the lowest energy at bond length 3.75\AA \hspace{0.1mm} is, \begin{align}
  \ket{\psi_s} &= \frac{1}{2}(\ket{1111111111111100000000000000} - \ket{1111111111110011000000000000} \nonumber\\
&- \ket{1111111111001100110000000000} + \ket{1111111111000011110000000000}). \nonumber
\end{align}

The stabilizer code for the above state is a [[28, 1, 2]] code and has the following logical operators, $\Bar{X} = Z_{16}Z_{18}$, $\Bar{Z} = -X_{10}X_{11}X_{14}X_{15}$, and the following stabilizers, $\mathcal{S}^{'} = \{ -Z_1, -Z_2, $ $-Z_3, -Z_4, -Z_5, -Z_6, -Z_7, -Z_8, -Z_9,$ $-Z_{10}Z_{15},$ $-Z_{11}Z_{15},$ $-Z_{12}Z_{17},$ $-Z_{13}Z_{17},$  $Z_{14}Z_{15},$  $Z_{16}Z_{17},$  $X_{10}X_{11}X_{12}X_{13}X_{14}X_{15}X_{16}X_{17},$ $Z_{18},$ $Z_{19},$ $Z_{20},$ $Z_{21},$ $Z_{22},$ $Z_{23},$ $Z_{24},$ $Z_{25},$ $Z_{26},$ $Z_{27},$ $Z_{28}\}$.

\subsubsection{Cr$_2$ molecule}
The stabilizer state, $\ket{\psi_s}$, with the lowest energy at bond length 5.05\AA \hspace{0.1mm} is,
\begin{align}
    \ket{\psi_s} &= \frac{1}{4}(
      \ket{111111111111000000000000000000000000} -\ket{111111111100000000001100000000000000}\nonumber \\
    &-\ket{111111110011000000000011000000000000} +\ket{111111110000000000001111000000000000}\nonumber \\
    &+\ket{111110101111000000000000000000010001} -\ket{111110101100000000001100000000010001}\nonumber \\
    &-\ket{111110100011000000000011000000010001} +\ket{111110100000000000001111000000010001}\nonumber \\
    &+\ket{111101011111000000000000000010001000} -\ket{111101011100000000001100000010001000}\nonumber \\
    &-\ket{111101010011000000000011000010001000} +\ket{111101010000000000001111000010001000}\nonumber \\
    &+\ket{111100001111000000000000000010011001} -\ket{111100001100000000001100000010011001}\nonumber \\
    &-\ket{111100000011000000000011000010011001} +\ket{111100000000000000001111000010011001}) \nonumber
\end{align}.

The stabilizer code for the above state is a [[36, 1, 2]] code and has the following logical operators, $\Bar{X} = Z_{22}Z_{24}$, $\Bar{Z} = -X_{11}X_{12}X_{21}X_{22}$, and the following stabilizers, $\mathcal{S}^{'} = \{-Z_1,$ $-Z_2,$ $ -Z_3,$ $-Z_4,$ $-Z_{5}Z_{33},$ $-Z_{6}Z_{36},$ $-Z_{7}Z_{33},$ $-Z_{8}Z_{36},$ $-Z_{9}Z_{24},$ $-Z_{10}Z_{24},$ $-Z_{11}Z_{22},$ $-Z_{12}Z_{22},$ $Z_{21}Z_{22},$ $Z_{23}Z_{24},$ $Z_{29}Z_{33},$ $Z_{31}Z_{36},$ $X_{6}X_{8}X_{32}X_{36},$  $X_{5}X_{7}X_{29}X_{33},$ $X_{9}X_{10}X_{23}X_{24},$ $Z_{13},$ $Z_{14},$ $Z_{15},$ $Z_{16},$ $Z_{17},$ $Z_{18},$ $Z_{19},$ $Z_{20},$ $Z_{25},$ $Z_{26},$ $Z_{27},$ $Z_{28},$ $Z_{30},$ $Z_{31},$ $Z_{34},$ $Z_{35}\}$.
\bigskip

\subsection{Example preparation of generalized stabilizer states}\label{app:ex_gen_code}
In this section we present some examples of generalized stabilizer states for the different molecules considered in this study that can be prepared using the error-detection codes.

\subsubsection{H$_4$ molecule}
The stabilizer state, $\ket{\psi_s}$, with the lowest energy at bond length 3.0\AA \hspace{0.1mm} is,
\begin{align}
    \ket{\psi_s} &= \frac{1}{2}(\ket{11110000} - \ket{11000011} - \ket{00111100} + \ket{00001111}). \nonumber
\end{align}

The logical states, $\Bar{\ket{\pm}}$, of the code for this state in Sec.~\ref{app:ex_code}-2., are $\ket{\alpha} = 1/\sqrt{2} (\ket{11110000} + \ket{00001111})$ and $\ket{\beta} = -1/\sqrt{2} (\ket{11000011} + \ket{00111100})$.
Thus, using the same code and universal fault-tolerant operations, we can prepare a generalized stabilizer state, $\ket{\psi} = a\ket{\alpha} + b\ket{\beta})$, where $a$, $b$ are real numbers and $a^2+b^2=1$.

\subsubsection{BH$_3$ molecule}
The logical states, $\Bar{\ket{\pm}}$, of code in Sec.~\ref{app:ex_code}-3., are 
\begin{align}
    \ket{\alpha} &= 1/\sqrt{2} (\ket{111111000000} + \ket{110000111100}  + \ket{110100111000} +\ket{111011000100}), \text{and }\nonumber \\
    \ket{\beta} &= -1/\sqrt{2} (\ket{111000110100} + \ket{110100111000} + \ket{111100110000} + \ket{110111001000}). \nonumber
\end{align} 
Thus, using the same code and universal fault-tolerant operations, we can prepare a generalized stabilizer state, $\ket{\psi} = a\ket{\alpha} + b\ket{\beta})$, where $a$, $b$ are real numbers and $a^2+b^2=1$.

\subsection{Circuit for preparing the generalized stabilizer states}\label{app:noisy_cir}
\begin{figure}[htbp!]
    \centering
    \includegraphics[width=0.99\columnwidth]{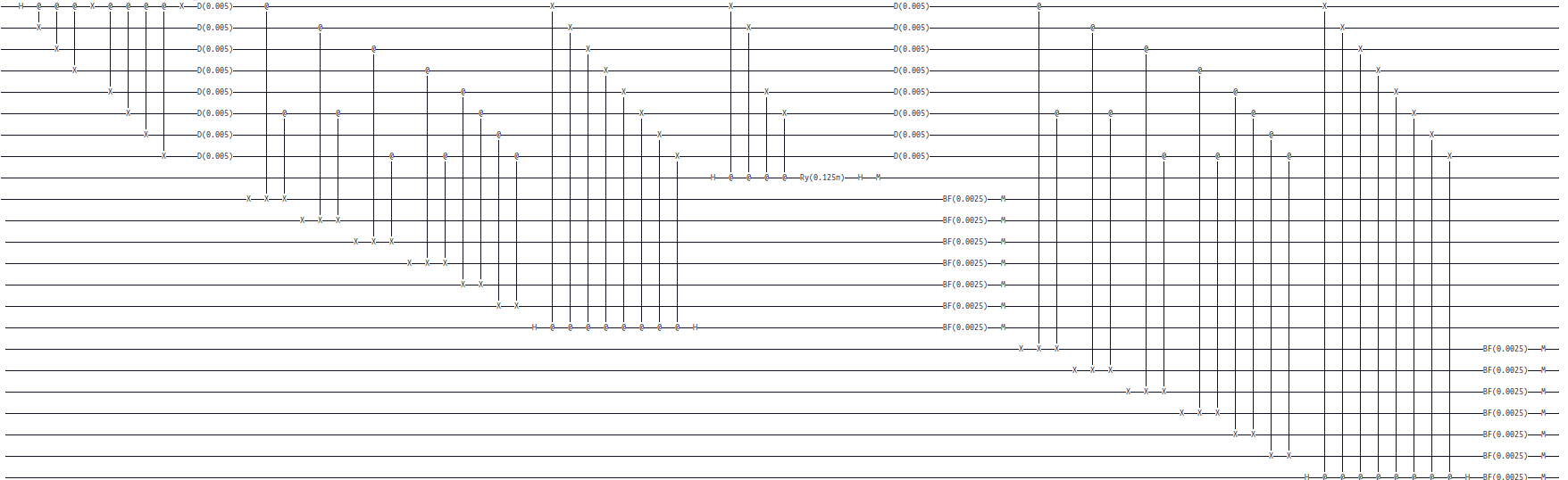} \\
    \caption{\label{fig:noisy_circuit} The circuit used for noisy preparation of a generalized stabilizer state for the case of H$_4$ molecule.} 
\end{figure}

\end{document}